# VIBRATIONAL INFRARED LIFETIME OF THE ANESTHETIC NITROUS OXIDE GAS IN SOLUTION

LOGAN CHIEFFO[†,‡], JASON J. AMSDEN[*,‡], JEFFREY SHATTUCK[†,‡], MI K. HONG[*,‡], LAWRENCE ZIEGLER[†,‡,#], SHYAMSUNDER ERRAMILLI[*,‡,$,#]

[†]Department of Chemistry, Boston University, 590 Commonwealth Avenue, Boston, MA-02215, USA
[*]Department of Physics, Boston University, 590 Commonwealth Avenue, Boston, MA-02215, USA
[$]Department of Biomedical Engineering, Boston University, 44 Cummington Street, Boston, MA-02169, USA
[‡]Center for Photonics, Boston University, 8 St. Mary's Street, Boston, MA-02169, USA

[#] To whom correspondence should be addressed

**Abstract.** The lifetime of the asymmetric fundamental stretching 2218 $cm^{-1}$ vibration of the anesthetic gas nitrous oxide ($N_2O$) dissolved in octanol and olive oil is reported. These solvents are model systems commonly used to assess anesthetic potency. Picosecond time-scale molecular dynamics simulations have suggested that protein dynamics or membrane dynamics play a role in the molecular mechanism of anesthetic action. Ultrafast infrared spectroscopy with 100 fs time resolution is an ideal tool to probe dynamics of anesthetic molecules on such timescales. Pump-probe studies at the peak of the vibrational band yield a lifetime of 55 ± 1 ps in olive oil and 52 ± 1 ps in octanol. The similarity of lifetimes suggests that energy relaxation of the anesthetic is determined primarily by the hydrophobic nature of the environment, consistent with models of anesthetic action. The results show that nitrous oxide is a good model system for probing anesthetic-solvent interactions using nonlinear infrared spectroscopy.

*Key words*: Anesthesia, ultrafast, laser, femtosecond, pump-probe, infrared spectroscopy, nitrous oxide, membranes

## 1. Introduction

The phenomenon of general anesthesia was discovered more than 150 years ago. Since then, the fundamental molecular nature of anesthetic action has remained a mystery. In molecular biophysics, the mechanism of many protein interactions, gene transcription and regulation, cell signaling and in fact most biochemical reactions are characterized by an extraordinary level of specificity. Anesthesia appears to be outside this dominant paradigm[1]. The list of general anesthetics includes a wide variety of chemically distinct classes of molecules. For example, halogenated hydrocarbons like halothane, certain *n*-alkanols, even noble gases like xenon, and nitrous oxide exhibit anesthetic potency. There is skepticism about whether one can explain in a unified picture the molecular mechanism of anesthetic action of such a wide range of anesthetic agents[1]. Femtosecond spectroscopy, imaging, and diffraction, combined with genetic engineering and mathematical modeling methods, have substantially increased our understanding of biomolecular structure and complex dynamics, and thus offer new probes for unraveling the molecular details of anesthetic action.

The first systematic studies on a large number of known anesthetic compounds were made almost exactly 100 years ago and showed that the solubility of anesthetics in olive oil correlates well with anesthetic potency. There was a general consensus that the binding site is a non-specific hydrophobic region in the lipid portion of cell membranes[2,3]. Within the past decade or two, long accepted notions regarding the nature of anesthetic action have been challenged [4,5]. Most recent thinking has converged upon the idea that the site of anesthetic action is a protein receptor [6-8] within a membrane protein or at the interface between a membrane protein and its surroundings. It is not known if there is a unique site of action or what the interaction mechanism may be. Computational methods of characterizing biomolecular structure and dynamics have yielded fresh insights [9-11]. Picosecond time-scale molecular dynamics simulations suggest that the anesthetic halothane influences the dynamics of lipid hydrocarbon tails, leading to a hypothesis that protein dynamics or membrane dynamics may play a role in regulating anesthetic action [12, 13]. Ultrafast infrared spectroscopy is a method that is uniquely suited to answering some of these questions. For example, ultrafast infrared spectroscopy can help determine whether the binding site is homogeneous or heterogeneous, and how the normal modes of anesthetic molecules may couple with the modes of proteins, lipids and/or water molecules near the binding site. Two singular aspects of anesthetic properties remain nearly universal: the reversal of the phenomenon of anesthesia due to hydrostatic pressure [14], and the correlation of potency with the solubility in olive oil[2]. Olive oil is a complex solvent containing triacylglycerols, free fatty acids like oleic acid, glycerol, pigments, aroma compounds, sterols, tocopherols, polyphenols, and many resinous components. Since olive oil is not a pure chemical compound, there has been a search for a simpler model system for correlates of anesthetic potency. Hydrophobicity appears to be the most relevant property, and in this regard, olive oil may be replaced by a number of other lipophilic solvents[15], such as lecithin or *n*-octanol. The solubility in *n*-octanol (i.e the gas-octanol partitition coefficient) and the octanol-water partition coefficient are both used to quantify the hydrophobic or lipophilic nature of the anesthetic molecule, and is often used as a good correlate of anesthetic potency[1]. In this regard, it is interesting that octanol is itself an anesthetic, so the solubility correlation holds for a range of general anesthetics, with the exception of octanol itself.

Nitrous oxide ($N_2O$) may be the oldest drug in continuous use in modern pharmacology, predating even aspirin. It has been in use since the phenomenon of anesthesia was discovered and continues to be widely employed in specific clinical applications. Although $N_2O$ is a weak anesthetic, its rapid uptake provides an advantage when used in combination with other more potent anesthetics. Use in combination with other anesthetics mitigates its deleterious effects [16](such as toxic side effects reported in children), and it is generally considered a safe anesthetic[17,18]. Recent appraisal of the complexity of action[1,19], and of its interaction with other anesthetics suggests that there is a multiplicity of binding sites, although there is strong evidence narrowing down the most likely target of action [16,19]. Clinical concentrations of nitrous oxide are known to inhibit NMDA-sensitive glutamate channels, suggesting that excitatory ligand-gated ion channels offer a pathway for anesthetic action [16,19]. The existence of heterogeneous binding sites for non-halogenated anesthetics is not established, and the mechanism of action at the molecular level has not been determined.

One advantage of nitrous oxide for studies of the molecular mechanism of general anesthetic action using IR spectroscopy is that it has simple, well-understood vibrational normal modes that are infrared active. Thus, binding site interactions can be studied without having to add any potentially perturbing labels such as fluorescent or spin tags. Nitrous oxide is a linear triatomic molecule. The infrared active modes include the symmetric stretching mode $\nu_1$ at 1285 cm$^{-1}$, the bending mode $\nu_2$ at 589 cm$^{-1}$ and a strong fundamental antisymmetric stretch $\nu_3$ centered at 2224 cm$^{-1}$, for $^{14}N_2^{16}O$ in the gas phase[20-23]. The antisymmetric stretch is especially attractive for IR spectroscopic studies because it falls in a window that is removed from the crowded part of the "fingerprint" infrared spectrum beyond about 5.5 microns in wavelength, where a large number of normal modes associated with proteins, nucleic acids and other biomolecules are present. The experimental gas phase dipole moment is 0.166 D[24], with the negative charge of the dipole at the oxygen end of the molecule ($NN^{+\delta}O^{-\delta}$). The sparse intramolecular vibrational level structure of a triatomic molecule like $N_2O$ makes ultrafast vibrational energy relaxation measurement a sensitive probe of solute-solvent interactions. Our report is the first application of ultrafast infrared spectroscopy to study the binding of nitrous oxide to model anesthetic targets.

**2. Materials and Methods**

*Sample preparation:* Octanol and olive oil were purchased from Sigma Chemicals and used without further purification. Nitrous oxide (99%) was supplied by Linde gas. The solvents were placed in a mixing chamber where they were pre-saturated with nitrous oxide. The design of the home-built steel pressure cell allowed for it to be coupled directly to the mixing chamber, without loss of nitrous oxide from the saturated solution during transfer to the sample cell. The cell was evacuated under vacuum prior to filling, and the equilibrated sample was sandwiched between calcium fluoride windows. The pathlength was set by a 100 µm teflon spacer. The same sealed cell was used for both static FTIR and ultrafast infrared measurements.

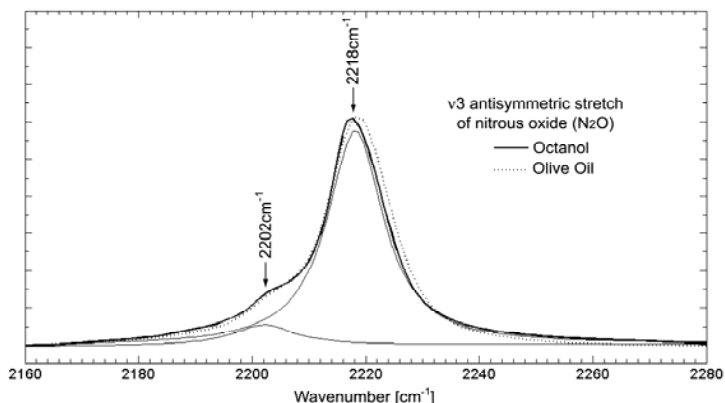

**Figure 1**. Static FTIR absorbance spectrum of nitrous oxide in the hydrophobic solvents olive oil (bold solid line) and octanol (dashed line) showing the antisymmetric stretch mode with a peak at 2218 cm$^{-1}$. The small band at 2202 cm$^{-1}$ is assigned to a hot band as described in the text. The thin solid lines show the two lorentzians that were fit to the FTIR of octanol.

*FTIR spectroscopy:* FTIR spectroscopy measurements were performed on nitrous oxide in octanol and olive oil to determine the absorption cross-section and spectral band shapes. Spectra were collected using a commercial FTIR spectrometer equipped with a liquid-nitrogen cooled mercury cadmium telluride (MCT) detector at a spectral resolution of 2 cm$^{-1}$. FTIR measurements were made before and after ultrafast measurements to check for loss of nitrous oxide due to leaks. The pressure of nitrous oxide studied ranged from $10^5$ Pa (1 bar) to $10^6$ Pa (10 bar), an order of magnitude below the pressure range that is known to induce pressure reversal of anesthesia. Figure 1 shows the FTIR measurements at 1 bar of nitrous oxide at room temperature (298K).

*Ultrafast IR studies:* Our femtosecond IR spectrometer system is based on optical parametric amplification and difference frequency generation in a silver gallium sulfide (AgGaS$_2$) crystal pumped by a commercial regenerative Ti:sapphire amplifier (Spectra Physics). A schematic of the ultrafast IR pump-probe system is shown in figure 2. Briefly, a 130-femtosecond mid-infrared laser pulse (2.8 µJ) is tuned to the N$_2$O ν$_3$ absorption band. A 1:2 reflective telescope enlarges and collimates the IR beam to provide a tighter focal spot size at the sample (100μm diameter). A 50-50 CaF$_2$ beamsplitter is then used to split the beam into the pump and probe arms. A pair of two inch retro-reflectors (Prosystem inc.), mounted on computer controlled translation stages with 10nm resolution (Melles Griot), provides the relative delay between the pump and probe. The beams are focused onto the sample and recollimated using a pair of unprotected gold coated off-axis parabolic mirrors (focal length = 101.6mm). After the sample, the probe beam is focused by a CaF$_2$ lens onto the entrance slit of a monochromator (Acton), equipped with a 150 groove/mm grating blazed at 4μm. The light exiting the monochromator is either passed through the exit slit and detected with an indium-antimony (InSb) detector (Electro-Optical Systems). Signals from the detectors are digitized by a National Instruments DAQ board.

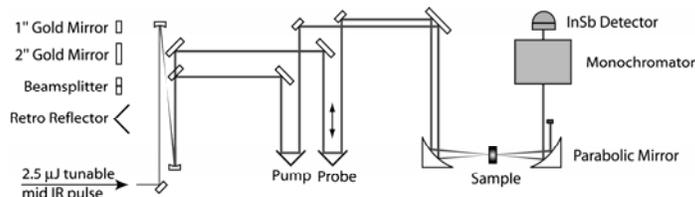

**Figure 2**. Schematic illustration of the pump-probe experiment. Mid-IR pulses enter the system and are enlarged by a one to two reflective telescope before being split into pump and probe arms. The timing between the pump and probe arms is set by a computer controlled delay stage. Parabolic mirrors are used to focus pump and probe into the sample volume. The probe beam is passed through a monochromator and is detected by an indium-antimony detector.

To aid in alignment, a portion of the 800nm beam is picked off before entering the optical parametric amplifier and is overlapped with the mid-IR beam before entering the experimental setup region. A ZnSe window with a broad band IR antireflection coating is used to overlap the IR and 800 nm beams. This laser system provides tunable 100-fs pulses of infrared radiation throughout the mid IR. The pulse width is checked by measuring the autocorrelation in a ZnGeP$_2$ crystal.

*Magic Angle Measurements:* Calcium fluoride holographic wire grid polarizers (Optometerics LLC) were place in both the pump and probe arm before the last turning mirror and sample. The pump polarizer was rotated 54.7 degrees with respect to the probe polarizer, to obtain magic angle measurements. All experiments were carried out at ambient room temperature.

### 3. Results and Discussion

*Static FTIR studies*: The static FTIR spectra of the $\nu_3$ antisymmetric stretch of nitrous oxide in both olive oil and octanol were fit with two Lorentzian functions (see figure 1). The fits of the nitrous oxide absorption band in both solvents gave very similar results. The dominant feature is the $\nu_3$ antisymmetric stretch fundamental centered at 2218cm$^{-1}$ in octanol (2219 cm$^{-1}$ in olive oil). The peak frequency is red shifted by ~ 6 cm$^{-1}$ relative to the gas phase frequency[23, 25]. This observation is consistent with the studies of Caughey and co-workers[26, 27]. When N$_2$O is in a polar hydrophilic environment such as water the peak frequency undergoes a blue shift to 2230 cm$^{-1}$ compared to the gas phase frequency of 2224 cm$^{-1}$. Detailed studies of nitrous oxide bound to the hydrophilic solvents and lyotropic lipid-water systems will be presented elsewhere. The smaller peak at 2202 cm$^{-1}$ in octanol (2203 cm$^{-1}$ in olive oil) is assigned to a "hot band". The hot band is due to excitation from the thermally populated $\nu_2$ bending mode to the $\nu_2 + \nu_3$ combination level. This peak is redshifted with respect to the $\nu_3$ fundamental peak transition by anharmonic coupling between the bending mode ($\nu_2$) and the antisymmetric stretch ($\nu_3$).

The extinction coefficients were determined in order to assess the suitability of N$_2$O for nonlinear ultrafast studies. A peak extinction coefficient of 1.5×10$^3$ M$^{-1}$ cm$^{-1}$ was determined for N$_2$O in both olive oil and octanol. The similarity of the FTIR spectra in the two solvents lends support to the notion that nitrous oxide occupies similar environments in both solvents. In particular, N$_2$O is associated with the hydrocarbon tails of the molecules, consistent with the hydrophobicity of the binding site being important to the mechanism of anesthetic action[1].

*Vibrational energy relaxation studies:* Vibrational energy relaxation measurements are shown in figure 3. Several points may be noted. (i) The size of the signal ranged from 10-25 mOD, depending on the concentration and pressure. Our experimental system is able to make measurements at the level of 10$^{-2}$ mOD, thus providing a large dynamic range. Therefore, prospects for analyzing more complex decay processes are excellent. (ii) Apart from the sharp spike near time $t$=0, which is due to sub-picosecond coherence effects, the magic angle data for $N_2O$ in olive oil and octanol out to $> 10^2$ ps can be fit to a single exponential decay curve. Using olive oil (figure 3a) as the solvent, the time constant was 55 ± 1 ps. Using octanol (figure 3b) as the solvent, the time constant was 52 ± 1 ps.

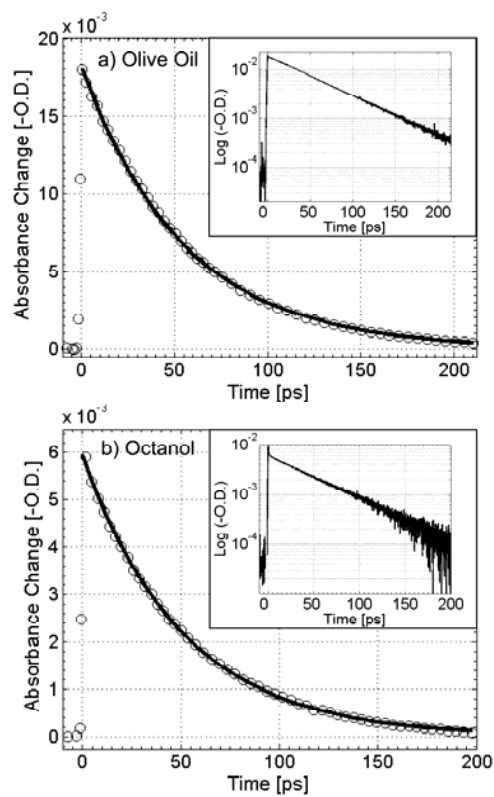

**Figure 3**. Vibrational energy lifetime of $N_2O$ (magic angle polarization) in (a) olive oil and (b) octanol. The circles are experimental points and the line is a fit to the data. Inset are semi-log plots of the experimental data. The data was fit to a single exponential with decay constant of 55±1ps and 52±1 ps for olive oil and octanol.

As stated before olive oil is a complex solvent. It is interesting that even in such a complex heterogeneous system, the vibrational energy relaxation data can be fit to a simple exponential. Vibrational energy relaxation can in general occur due to the coupling to other intramolecular modes, anharmonic effects, coupling to high frequency solvent modes or low frequency solvent modes that constitute the thermal bath[28]. The sparseness of the nitrous oxide vibrational spectrum and the lack of other intramolecular high frequency modes near $v_3$ suggests that the dominant contribution to the observed relaxation should be due to interaction with the solvent as compared to an intramolecular vibrational energy relaxation pathway[28]. The dipole moment of $N_2O$ is smaller than that of water. Relaxation in polar solvents like water is expected to be faster than in hydrophobic solvents due to the more significant dipole-induced dipole forces. $T_1$ is nearly the same for $N_2O$ in both octanol and olive oil. This is noteworthy because the two solvents differ in viscosity by more than an order of magnitude, with $\eta$ = 84 cP (8.4 Pa s) for olive oil, and 7.3 cP (0.73 Pa s) for octanol at 298 K. On the other hand, the hydrocarbon tails of the two systems are both hydrophobic. This observation suggests that even on ultrafast time-scales, the dynamics of anesthetic-solvent interaction is determined not so much by molecular level details or due to any bond formation between the solute and solvent, but rather a generic property like the hydrophobicity, known to be a strong correlate of anesthetic action.

*Magic Angle Studies:* Rotational motion could contribute to the pump-probe decays at the earliest times for small molecules like $N_2O$ in liquids at room temperature. Magic angle measurements substantially reduce rotational contributions, although the interpretation of magic angle measurements in ultrafast studies has pitfalls[29]. Magic angle studies suggest that rotational effects affect the relaxation on fast time scales (< 2 ps) but are not important on the long time scales that characterize the energy relaxation observations presented here.

## 4. Conclusions

Our report is the first application of femtosecond infrared pump-probe spectroscopy to study the energy relaxation of nitrous oxide in the model hydrophobic solvents, octanol and olive oil. The similarity of the observed lifetimes of $N_2O$ in the two solvents suggests that energy relaxation of the anesthetic is determined primarily by the hydrophobic nature of the environment, consistent with models of anesthetic action. Nitrous oxide is a good model system for probing anesthetic-solvent interactions using nonlinear infrared spectroscopy.

**Acknowledgement.** Support is gratefully acknowledged from the U.S. National Science Foundation and the Department of Defense. We thank Sarah Keller and Christopher Rella for advice on preliminary experiments. J.A. was supported by an NIH grant (GM069969). S.E. thanks the organizers of the 2006 Asia-Pacific Workshop on Biological Physics.